\documentclass{ccjnl}

\usepackage{bm}

\ccjnlset{
  title = {Low-Complexity Channel Estimation for RIS-Assisted ISAC System},
  corr-email = {chenz.scut@gmail.com},
}

\author*[addr1]{Zhen Chen}[style=chinese]
\author[addr2]{Jianqing Li}[style=chinese]
\author[addr3]{Haijun Zhang}[style=chinese]
\author[addr4]{Wei Zhang}[style=chinese]

\address[addr1]{The Institute of Microelectronics, University of Macau, Macau}
\address[addr2]{School of Computer Science and Engineering, Macau University of Science and Technology, Macau, China}
\address[addr3]{Beijing Engineering and Technology Research Center for Convergence Networks and Ubiquitous Services, University of Science and Technology Beijing, Beijing, China}
\address[addr4]{Henan Academy of Science Applied Physics Institute Co.,Ltd, Henan Zhengzhou, China}

\begin{document}
\maketitle

\begin{abstract}
Integrated sensing and communication (ISAC), assisted by reconfigurable intelligent surface (RIS) has emerged as a breakthrough technology to improve the capacity and reliability of 6G wireless network. However, a significant challenge in RIS-ISAC systems is the acquisition of channel state information (CSI), largely due to co-channel interference, which hinders meeting the required reliability standards. To address this issue, a minimax-concave penalty (MCP)-based CSI refinement scheme is proposed. This approach utilizes an element-grouping strategy to jointly estimate the ISAC channel and the RIS phase shift matrix. Unlike previous methods, our scheme exploits the inherent sparsity in RIS-assisted ISAC channels to reduce training overhead, and the near-optimal solution is derived for our studied RIS-ISAC scheme. The effectiveness of the element-grouping strategy is validated through simulation experiments, demonstrating superior channel estimation results when compared to existing benchmarks.
\end{abstract}

\section{introduction}
\label{s1}
The integrated sensing and communication (ISAC) technology has been foreseen as a promising candidate  for supporting emerging 6G wireless networks, enabling advancements in ubiquitous sensing, intelligence, and connectivity, etc \cite{10118701,10135096,10143420}. By integrating both sensing and communication (SAC) functionalities on the same hardware platform and using a single frequency band, ISAC enhances spectrum efficiency while significantly reducing hardware infrastructure costs. To support simultaneous information transmission and target detection, numerous works have devoted the efforts to the design of ISAC system, including power optimization \cite{10325366}, degrees of freedom estimation \cite{9906898} and waveform design \cite{9724187}. According to the third Generation Partnership Project (3GPP), vehicle-to-everything (V2X) communications will facilitate efficient information exchange between sensing and communication, which highlighting  a substantial potential market for a wide range of intelligent vehicle applications \cite{8809218}.

Among these literatures, an accurate channel state information (CSI)  is crucial for exsiting ISAC design, because the promising gains brought by ISAC frameworks depend on CSI. However, the existing scheme for ISAC channel  is practically difficult to acquire due to  the assumption of identical dominant paths. Thus,  the corresponding  estimation procedure is confined solely to that particular scenario. To fulfill the increasing demand for communication and sensing performance, several studies have focused on the fundamental model of ISAC channels \cite{10225688}. Unlike traditional communication channels, ISAC channels  involve the radio propagation paths from transmitters to facilitate communication between devices, and echo propagation paths for sensing applications.
For instance,  Cheng et al. \cite{9830717}  develop functional ISAC of Internet of Vehicles (IoV) networks to enhance sensing and/or communication channels, regardless of the sensor type. In \cite{10448481}, the non-cooperative moving targets are localized and the associated propagation paths are excluded from the SAC channel. To  tackles the challenge of general memoryless point-to-point channel in  vehicular network scenarios, an correlated sensing scheme was investigated to provide a significant gain, while maintaining sensing and communication performance \cite{9787809}.

However, the significant distance between the vehicle and the cloud server, coupled with limited data link capacity  \cite{9622178}.  This issue hinders the efficient transmission of data required, affecting overall performance and user experience. To address this issue, the reconfigurable intelligent surface (RIS), a revolutionary and energy-efficient technique, has garnered considerable attention due to its potential in sensing and communication channel estimation \cite{10218356,9521836,10186271}. Specifically, the phase-shift matrix of the RIS can be adjusted intelligently for colocated MIMO radar, and respectively, communication channel to enhance the communication and detection performance \cite{9361184}. Based on these concepts, RIS was  successfully integrated into  ISAC system to enhance the surrounding  vehicle communication  environment intelligently \cite{10218356, 10186271}. In \cite{10218356}, the RIS was investigated to alleviate the environments in  severe path loss case that can improve target detection of ISAC system. The authors in
\cite{10186271} consider RIS-aid beampattern design to improve the capabilities of a MIMO-ISAC system. Since RIS has shown great coverage advantages and provide extra reflecting links, it is expected to provide high sensing accuracy while guaranteeing the communication performance. To further reduce the pilot signal overhead, the approach involves leveraging data-driven deep-learning (DL) techniques for channel estimation was investigated to reduce pilot signal overhead and achieving high estimation accuracy \cite{10109100}. Moreover, the spatial-delay domain coarse channel estimation can be conducted with ray tracing to characterize communication and sensing channels \cite{10286864}. In \cite{10078840}, the  forward and backward scattering was investigated, where the correlations between sensing and communication channels was exploited to obtain the information of the scatterer. More recently, an   OTFS channel estimation was developed by expoiting deep learning strategy in complex scenarios \cite{10349836}. In \cite{10128162}, a pulse-based ISAC receiver for multi-path channels was designed to facilitate joint data recovery and ranging estimation. Recently, an iterative NOMA-ISAC scheme for channel state information (CSI)-based sensing was proposed; this scheme separates sensing and communication signals in either the time or frequency domains \cite{10082967}. However,  ensuring and preserving the power difference criteria necessary for implementing Successive Interference Cancellation (SIC) proves challenging, particularly in scenarios involving mobile targets or users. Recently, a flexibly tunable RIS is fabricated to extending the coverage and enhancing the throughput \cite{9999288}. In addition, multi-scenario broadband channel measurements and modeling were conducted for RIS-assisted communications, which cover distance and angle domains \cite{10319362}. In general, the ISAC channel estimation research of ISAC is still in its infancy and has become a hot topic in 6G IoV communication research.

Motivated by these studies, we intend to employ RIS to address the sensing and communication channel in ISAC systems, which is still an open issue. Different from the above studies, the RIS-assisted ISAC channel estimation is proposed to provide a better trade-off between communication and radar performance.  To reduce the training overhead, the weighted minimax-concave penalty (MCP) with sparsity inherent in ISAC channels is exploited to guarantee the convexity of the sparsity-regularized cost function. Then, an iterative ISAC channel estimation with high accuracy and low complexity
are proposed, where a joint SAC channel and RIS phase shift matrix design is developed by enforcing an  element-grouping  constraint. Extensive simulations are conducted to support the connected and automated vehicles.
Our main contributions are summarized as follows:

\begin{itemize}
	\item We propose a low-complexity channel estimation approach using a weighted minimax-concave penalty (MCP) technique, which effectively exploits the sparsity inherent in RIS-assisted ISAC systems. This novel method significantly reduces training overhead while maintaining high estimation accuracy, setting it apart from traditional channel estimation techniques.
	\item To further optimize the channel estimation process, we introduce an element-grouping strategy that shares reflection elements across groups. This reduces the dimensionality and complexity of the estimation problem, providing a practical solution for large-scale RIS arrays in ISAC networks.
	\item The proposed method demonstrates superior performance compared to existing benchmarks, such as the LS method, in terms of normalized mean square error (NMSE) under various SNR conditions and different network configurations. The proposed approach offers a better trade-off between sensing and communication performance in RIS-assisted systems.
\end{itemize}

The remaining parts of this paper are organized as follows. The ISAC channel model is formulated in Section \ref{section2}. An MCP-based channel estimation approach is designed in Section \ref{section3}. Section \ref{section4} designs the alternating iterative algorithm for ISAC channel estimation. Simulation results are
shown in Section \ref{SIMULATION-RESULTS}. Section \ref{section6} concludes this article.

$\emph{Notation}$: $\textbf{\emph{a}}$, $\emph{a}$ and A represent a vector and a scalar, and a set,
respectively; $\mathbb{C}$ denotes the field of complex numbers; $\|\textbf{\emph{a}}\|_2$ denotes the 2-norm of vector $\textbf{\emph{a}}$;  $\mathbb{R}^n$ stands for the set of $n$-vector real numbers; The operator $\mathrm{diag}(\mathrm{A})$
forms a column vector out of its diagonal elements; $\mathbb{E}[a]$ means the expectation  operator  of $\emph{a}$.

\begin{figure}[!t]
	\includegraphics[width=3.5in]{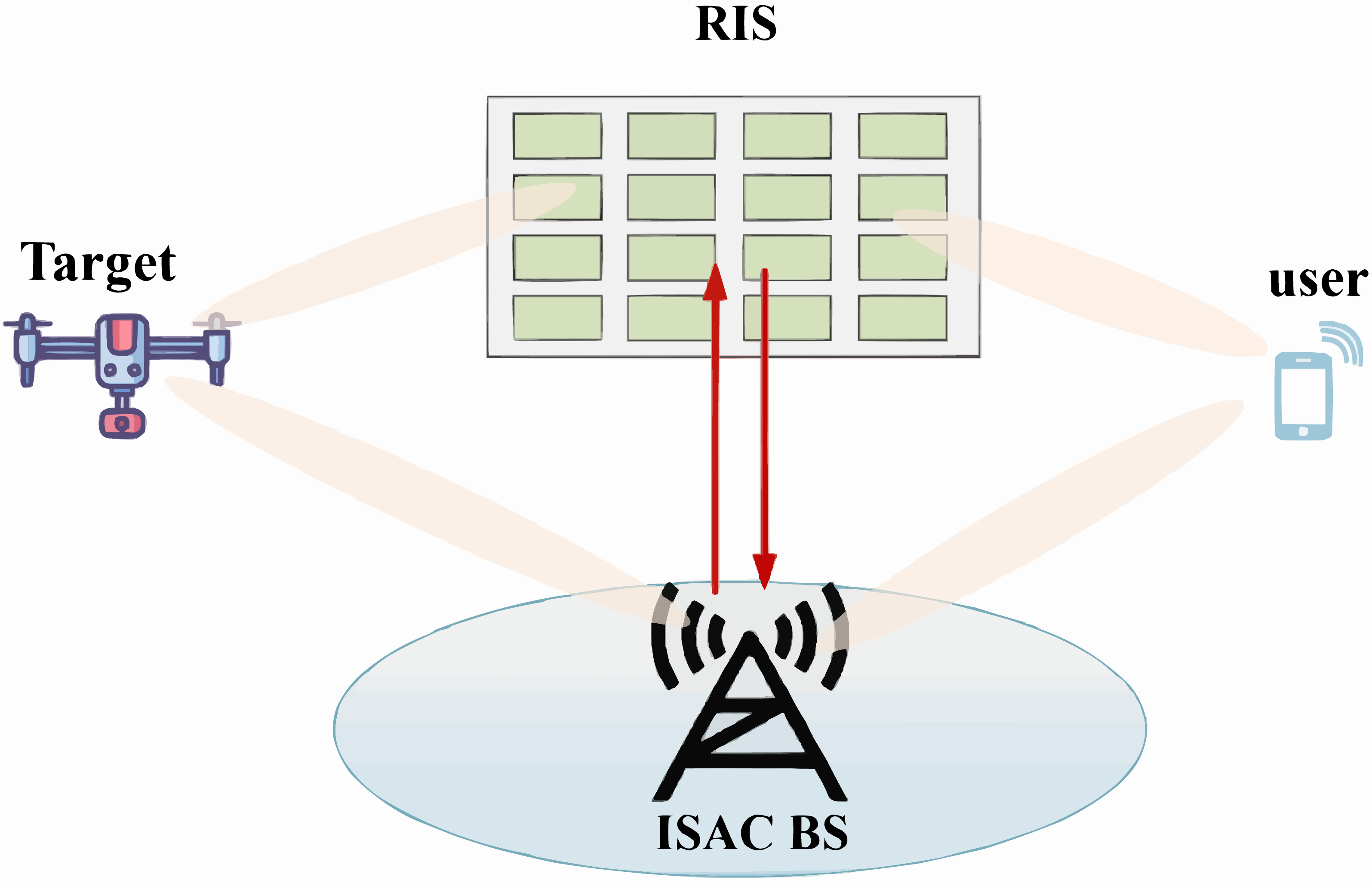}
	\caption{Illustration of the considered RIS-assisted ISAC system.}
	\label{Fig.1}
\end{figure}

\section{Problem Formulation}\label{section2}
Considering an RIS-assisted ISAC system with an full-duplex (FD) base station (BS), the BS and the RIS respectively employ
the $M$-antenna and the $L$-element uniform planer array (UPA) to simultaneously serve a target and a user. The ISAC BS is equipped with $M$ transmit antennas and $M$ receive antenna to  detect the target, while the user tries to communicate with BS via RIS reflection. 
Since system performance depend on the RIS deployment location, the RIS is deployed to be close 
the BS to promote the benefit of the RIS reflection \cite{9997576}. For senesing  processing, the ISAC BS transmits the radar probing waveforms to detect the target, and then the echo signals will be received through the direct BS$\rightarrow$target$\rightarrow$BS link $\emph{\textbf{d}}\in\mathbb{C}^{M\times 1}$ and reflected BS$\rightarrow$target$\rightarrow$RIS$\rightarrow$BS links  $\emph{\textbf{A}}\in\mathbb{C}^{L\times M}$. For communication processing,  the signals are received by combining the direct user$\rightarrow$BS and reflected user$\rightarrow$RIS$\rightarrow$BS channels. We also define the communication channel from user to BS links as $\emph{\textbf{f}}\in\mathbb{C}^{M\times 1}$, while the channel from RIS to BS links are defined as $\emph{\textbf{H}}\in\mathbb{C}^{L\times M}$  and $\emph{\textbf{w}}\in\mathbb{C}^{L\times1}$ denotes the channel between the RIS and the user.  In addition, the self-interference (SI) channel $\emph{\textbf{h}}_{si}\in\mathbb{C}^{M\times1}$ is employed to the ISAC BS, due to the FD mode. Multipath components with the same delay $\tau_l$ is considered in each cluster $l$. Then, the geometric channel model is employed as
\begin{equation}
	\begin{aligned}
		\emph{\textbf{h}}_{T,d}= \sqrt{\frac{M}{\rho_T}}\sum\limits_{l=1}^L\beta_Lp(dT_s-\tau_l)\emph{\textbf{a}}(\theta_l,\phi_l),
	\end{aligned}
\end{equation}
where $L$ represents  the total number of clusters. The variables  $\theta_l$ and $\phi_l$ denote the azimuth and elevation angles of arrival (AoA) for cluster $l$, respectively. Moreover, $\rho_T$ is the propagation pathloss at transmitter-receiver link, and $p(\tau)$ denotes a pulse shaping function at $\tau$ seconds.

Let $\emph{\textbf{s}}$ be the sensing signal sent from the ISAC BS, while the pilot symbol $\mathbf{z}$ is used at wireless transmission. Thus, the received signal at the ISAC BS can be written as
\begin{equation}\label{eq:001}
	\begin{aligned}
		\hat{\emph{y}}_{t}= & \underbrace{\left(\emph{\textbf{d}}^{H}+\emph{\textbf{w}}^H\operatorname{diag}\{\bm{\theta}_t\} \emph{\textbf{A}}\right)\emph{\textbf{s}}}_{\text {Sensing signal }}+\underbrace{\emph{\textbf{h}}_{si}^{H}\emph{\textbf{s}}}_{\text {Residual SI }} \\
		&\quad +\underbrace{\left(\emph{\textbf{f}}^H+\emph{\textbf{w}}^H\operatorname{diag}\{\bm{\theta}_t\}\emph{\textbf{H}}\right)\mathbf{z}}_{\text { Communication signal }}+\emph{n}_{t},
	\end{aligned}
\end{equation}
where $\bm{\theta}_t=[\rho_t e^{\jmath \varphi_{t, 1}}, \rho_t e^{\jmath \varphi_{t, 2}}, \ldots, \rho_t e^{\jmath \varphi_{t, L}}]^H$ denotes
the RIS phase-shift vector. $\varphi_{t, l} \in[0,2 \pi)$ with $l \in \mathcal{N}^L$ is the phase-shift of the $l$-th RIS element. 
$\emph{n}_{t}$ denotes received noise that follows $\mathcal{C N}\left(0, \sigma_0^2\right)$ with the noise power $\sigma_0^2$.

Inspired by the propagation environment of ISAC scenario  \cite{9997576,9736621}, the slow-varying channel is considered to the sensing and communication channels for the ISAC system. Hence,  the residual SI term in \eqref{eq:001} can be achieved before estimating ISAC channels, where $\emph{\textbf{h}}_{si}$ can be estimated by employing the an LS approach. With straightforward derivation in \eqref{eq:001}, we can derive  $\emph{\textbf{w}}^H \operatorname{diag}\left\{\bm{\theta}_t\right\}=\bm{\theta}_t^H\operatorname{diag}\{\emph{\textbf{w}}\}$. By utilizing the RIS, the BS$\rightarrow$target$\rightarrow$RIS$\rightarrow$BS and VE$\rightarrow$RIS$\rightarrow$BS links can be written as $\emph{\textbf{G}}_{s}=\operatorname{diag}\left\{\emph{\textbf{w}}\right\}\emph{\textbf{A}} \in \mathbb{C}^{L\times M}$ and $\emph{\textbf{G}}_{\emph{c}}=\operatorname{diag}\{\emph{\textbf{w}}\}\emph{\textbf{H}}\in \mathbb{C}^{L \times M}$, respectively. Let $\emph{y}_{t}=\hat{\emph{y}}_{t}-\emph{\textbf{h}}_{si}^{H}\emph{\textbf{s}}$, the received signal in \eqref{eq:001} is generated as
\begin{equation}\label{eq:002}
	\emph{y}_{t}=\underbrace{\left(\emph{\textbf{d}}^{H}+\bm{\theta}_t^H\emph{\textbf{G}}_{\emph{s}}\right) \emph{\textbf{s}}}_{\text {Sensing signal }}+\underbrace{\left(\emph{\textbf{f}}^H+\bm{\theta}_t^H\emph{\textbf{G}}_{\emph{c}}\right) \mathbf{z}}_{\text {UL communication signal}}+\emph{n}_{t}.
\end{equation}

By denoting  $\tilde{\bm{\theta}}_t= [1\; \bm{\theta}_t^H]^H$, $\tilde{\emph{\textbf{G}}}_{\emph{s}}=[\emph{\textbf{d}}^{H};\emph{\textbf{G}}_{\emph{s}}]$ and $\tilde{\emph{\textbf{G}}}_{\emph{c}}=[\emph{\textbf{f}}^{H};\emph{\textbf{G}}_{\emph{c}}]$, the received signal \eqref{eq:002} can be written in a compact form as
\begin{equation}\label{eq:003}
	\begin{aligned}
		\emph{y}_{t}&=\underbrace{\tilde{\bm{\theta}}_t^{H}\tilde{\emph{\textbf{G}}}_{\emph{s}}\emph{\textbf{s}}}_{\text {Sensing signal }}+\underbrace{\tilde{\bm{\theta}}_t^H\tilde{\emph{\textbf{G}}}_{\emph{c}}\mathbf{z}}_{\text {UL communication signal}}+\emph{n}_{t}.
	\end{aligned}
\end{equation}

During the channel estimation processing, the ISAC BS transmits $T$ pilot symbols that are reflected by the RIS with the reflecting elements $\tilde{\bm{\theta}}_t$ for $t = 1, 2,...,T$. After $T$ time slots of pilot transmission, the  received signal is reformulated in a normalized form as
\begin{equation}\label{eq:004}
	\begin{aligned}
		\boldsymbol{y}&=\frac{1}{\sqrt{P}}[\emph{y}_{1},\emph{y}_{2},...,\emph{y}_{T}]^T
		&=\bm{\Theta}^T\tilde{\emph{\textbf{G}}}_{\emph{s}}\emph{\textbf{s}}+\bm{\Theta}^T\tilde{\emph{\textbf{G}}}_{\emph{c}}\mathbf{z}+\emph{\textbf{n}},
	\end{aligned}
\end{equation}
where $\bm{\Theta}= [\tilde{\bm{\theta}}_1, \tilde{\bm{\theta}}_2,..., \tilde{\bm{\theta}}_T]\in\mathbb{C}^{LM\times T}$ denotes  the reflection matrix of the RIS, and noise $\emph{\textbf{n}}=[\emph{n}_{1},\emph{n}_{2},...,\emph{n}_{T}]^T$ that satisfy $\emph{\textbf{n}}\sim\mathcal{CN}(0, \gamma^{-1}\emph{\textbf{I}}_T)$ with the SNR $\gamma=P/\sigma_0^2$, $P$ is the transmit power.

\begin{figure*}[!t]
	\centering
	\includegraphics[width=0.78\textwidth]{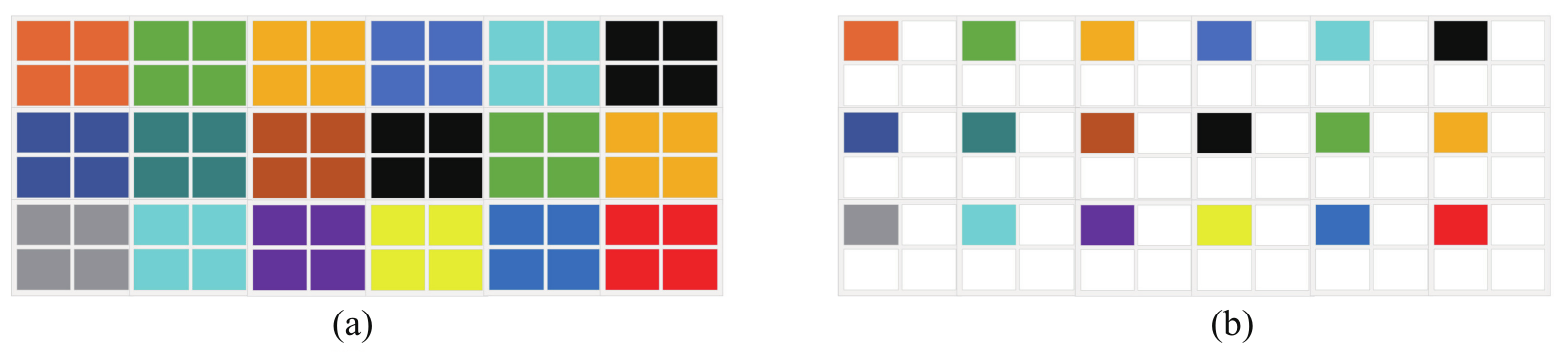}
	\caption{The element-grouping strategy for the training reflection pattern.}
	\label{Fig.2}
\end{figure*}

To tackle the enormous pilot overhead, we adopt the reflection matrix with element-grouping \cite{9530267}. As illustrated in Figure \ref{Fig.2}, the RIS elements can be divided into some subgroups that satisfies $L_Q=L/Q$, where  $Q$ adjacent elements is used for each subgroup to share the same reflecting elements. For the sake of fairness, the element-puncturing used for superresolution, or CSI extrapolation \cite{9621762}, is  illustrated in Figure \ref{Fig.2}(b), where one element from each element-grouping is chosen  for the training reflection.   
Let $\mathcal{S}_l$ be the $l$ th subgroup of RIS reflecting elements, we have $\left|\mathcal{S}_l\right|=Q$ for element-grouping and $\left|\mathcal{S}_l\right|=1$ for element-puncturing. The reflecting matrix of RIS with the subgroup-shared reflection $L_Q$ is reformulated as
\begin{equation}
	\bm{\Theta}^T=\bm{\Psi}\bm{\Omega},
\end{equation}
where $\bm{\Psi}\in \mathbb{C}^{M \times L_Q}$ denotes the reflection pattern, $\bm{\Omega}\in \mathbb{R}^{N_{a}\times N_{b}}$ is the subgroup mapping matrix that satisfies $\bm{\Omega}\bm{\Omega}^T=\left|\mathcal{S}_l\right| \emph{\textbf{I}}_{L_Q}$. Without loss of generality, if $\emph{m}\in\mathcal{S}_l$, we have $[\bm{\Omega}]_{\emph{l,m}}=1$, and $[\bm{\Omega}]_{\emph{l,m}}=0$ if $\emph{m} \notin \mathcal{S}_l$ for $l=1,2, \cdots, L_Q$. For the reflection pattern $\bm{\Psi}$, an unit modulus orthogonal matrix with $M=L_Q$ is exploited to satisfy $\bm{\Psi}^H\bm{\Psi}=\bm{\Psi\Psi}^H=L_Q \bm{I}_{L_Q}$ and $\left|[\bm{\Psi}]_{\emph{m,n}}\right|=1$.

Accordingly, the least square (LS) is exploited to the Eq.~\eqref{eq:004}, which is given by 
\begin{equation}\label{eq:006}
	\boldsymbol{r}=\left(\boldsymbol{\Psi}^H \boldsymbol{\Psi}\right)^{-1} \boldsymbol{\Psi}^H \boldsymbol{y}=\boldsymbol{\Omega} \tilde{\emph{\textbf{G}}}_{\emph{s}}\emph{\textbf{s}}+\boldsymbol{\Omega}\tilde{\emph{\textbf{G}}}_{\emph{c}}\mathbf{z}+\emph{\textbf{e}},
\end{equation}
where $\emph{\textbf{e}}=\left(\boldsymbol{\Psi}^H \boldsymbol{\Psi}\right)^{-1} \bm{\Psi}^H \emph{\textbf{n}}\sim \mathcal{C N}\left(\mathbf{0}, \gamma_{\emph{e}}^{-1} \bm{{I}}_{L_Q}\right)$ and $\gamma_{\emph{e}}=$ $L_Q\gamma$ represents the SNR level  by performing the training symbols  $L_Q$. By denoting $\emph{\textbf{S}}=\emph{\textbf{s}}^H\otimes\bm{\Omega}$ and $\emph{\textbf{C}}=\mathbf{z}^H\otimes\bm{\Omega}$, the observed channel is given by
\begin{equation}\label{eq:006x}
	\begin{aligned}
		\bm{r}&=\left(\emph{\textbf{s}}^H\otimes\bm{\Omega}\right)\emph{\textbf{g}}_{\emph{s}}+\left(\mathbf{z}^H\otimes\bm{\Omega}\right)\emph{\textbf{g}}_{\emph{c}}+\emph{\textbf{e}}\\
		&=\emph{\textbf{S}}\emph{\textbf{g}}_{\emph{s}}+\emph{\textbf{C}}\emph{\textbf{g}}_{\emph{c}}+\emph{\textbf{e}},
	\end{aligned}
\end{equation}
where $\emph{\textbf{g}}_{\emph{s}}=\mathrm{vec}(\tilde{\emph{\textbf{G}}}_{\emph{s}})$ and $\emph{\textbf{g}}_{\emph{c}}=\mathrm{vec}(\tilde{\emph{\textbf{G}}}_{\emph{c}})$.
Obviously, the received sensing and communication signals interfere with each other, which leads to difficult to decouple at the ISAC BS. As such, it is  challenging to find a feasible solution  in \eqref{eq:006}. To address this challenge, we develop a weighted  MCP-based scheme to estimate the sensing and communication channels, $\{\emph{\textbf{g}}_{\emph{s}}, \emph{\textbf{g}}_{\emph{c}}\}$, for the RIS-assisted ISAC system.

\section{Proposed Minimax-Concave Penalty based Estimation Approach}\label{section3}
In order to obtain a feasible solution, a practical channel estimation approach based on minimax-concave penalty is exploited.  In the following, a  minimax-concave penalty based estimation framework is then developed, which is formulated with maximum a posterior (MAP) problem

\begin{equation}\label{eq:007}
	\begin{aligned}
		 \left(\emph{\textbf{g}}_{\emph{s}}, \emph{\textbf{g}}_{\emph{c}}, \gamma_{\emph{e}}\right)&=\operatorname{argmax} \log P\left(\boldsymbol{r}\mid \emph{\textbf{g}}_{\emph{s}}, \emph{\textbf{g}}_{\emph{c}}, \gamma_{\emph{e}}\right) P\left(\emph{\textbf{g}}_{\emph{s}}, \emph{\textbf{g}}_{\emph{c}}, \gamma_{\emph{e}}\right) \\
		& =\operatorname{argmax} \log P\left(\boldsymbol{r}\mid\emph{\textbf{g}}_{\emph{s}}, \emph{\textbf{g}}_{\emph{c}}, \gamma_{\emph{e}}\right)\\
		&\qquad\qquad\quad+\log P(\emph{\textbf{g}}_{\emph{s}},\emph{\textbf{g}}_{\emph{c}})+\log P\left(\gamma_{\emph{e}}\right) \\
		& =\operatorname{argmax} \log P\left(\boldsymbol{r}\mid\emph{\textbf{g}}_{\emph{s}}, \emph{\textbf{g}}_{\emph{c}}, \gamma_{\emph{e}}\right)\\
		&\qquad\qquad\quad+\log P(\emph{\textbf{g}}_{\emph{s}})+\log P(\emph{\textbf{g}}_{\emph{c}}),
	\end{aligned}
\end{equation}

\noindent where $\emph{\textbf{g}}_{\emph{s}}$ and $\emph{\textbf{g}}_{\emph{c}}$ are the sensing channel and communication channel, respectively.  The variance $\gamma_{\emph{e}}$ represents independent with each other that satisfying $P\left(\gamma_{\emph{e}}\right) \propto 1$,
it is apparent that $\gamma_{\emph{e}}$ has equal probability within the range $[0, \infty)$.

The likelihood term $P\left(\boldsymbol{r}\mid\emph{\textbf{g}}_{\emph{s}}, \emph{\textbf{g}}_{\emph{c}}, \gamma_{\emph{e}}\right)$ is characterized by the Gaussian distribution with variance $\gamma_{\emph{e}}$. Note that the communication channel $\emph{g}_{c,i}=0$  corrupted with AWGN. For communication channel corrupted with impulse noise, $\emph{g}_{c,i} = \emph{r}_i- \emph{g}_{s,i}- \emph{e}_{i}$ is affected by the clipping of the pilot signal value range, which ensures that the corrupted pilot signal $\emph{r}_i$ always remains within the range [$I_{\mathrm{min}}; I_{\mathrm{max}}$], regardless of the actual dynamic range of the random valued impulse noise. This turns out that the distribution of $\emph{g}_{c,i}=0$ will not be a multimodal/uniform distribution. Instead, the $\emph{g}_{c,i}=0$ follows  the Laplcian distribution, which lead to  $P(\emph{\textbf{g}}_{\emph{c}})$ as an independently and identically distributed (i.i.d.)  zero-mean Laplacian distribution $P(\emph{g}_{\emph{c},i}) = \frac{1}{2h}e^{-|\emph{g}_{c,i}|/h}$. Then, the MAP estimator of Eq.~\eqref{eq:007} can be characterized as
\begin{equation}\label{eq:008}
	\begin{aligned}
		\left(\emph{\textbf{g}}_{\emph{s}}, \emph{\textbf{g}}_{\emph{c}}, \gamma_{\emph{e}}\right)= & \underset{\emph{\textbf{g}}_{\emph{s}}, \emph{\textbf{g}}_{\emph{c}}, \gamma_{\emph{e}}}{\operatorname{argmin}} \frac{1}{2 \gamma_{\emph{e}}^2}\|\boldsymbol{r}-\emph{\textbf{S}}\emph{\textbf{g}}_{\emph{s}}- \emph{\textbf{C}}\emph{\textbf{g}}_{\emph{c}}\|_2^2+\lambda\|\emph{\textbf{g}}_{c}\|_1 \\
		&+\eta J(\emph{\textbf{g}}_{\emph{s}})+T\log \gamma_{\emph{e}},
	\end{aligned}
\end{equation}
where $\lambda = \frac{1}{h}$, $P(\emph{g}_c)\propto \frac{1}{c} \mathrm{exp}(J(\emph{g}_s))$ is set as a Gibbs distribution, and $h$ is the standard deviation of $\emph{g}_{c,i}$. The solution of the components $\emph{\textbf{g}}_{c}$ can be obtained by employing the soft-thresholding strategy.
However, for soft-thresholding methods,  selecting an appropriate regularization parameter  $\lambda$ often is challenging, as it depends on the noise level associated with impulse noise.

To tackles this challenge, the Jeffreys' prior \cite{1973Bayesian}, the scale factor $\beta_i$ is employed to establish a general forms., $\emph{g}_{\emph{c},i}=\beta_i\emph{h}_{i}$, where each $\emph{g}_{\emph{c},i}$ satisfies independently identically distribution (iid). Let $\emph{h}_i$ and $\beta_i$ be independent of each other, the  modeling with respect to $\emph{g}_{\emph{c},i}$ can be directly expressed as
\begin{equation}\label{eq:009}
	\begin{aligned}
		&P(\emph{\textbf{g}}_{\emph{c}})=\prod_i P\left(\emph{g}_{\emph{c},i}\right), \\
		&P\left(\emph{g}_{c,i}\right) =\int_0^{\infty} P\left(\emph{g}_{c,i} \mid \emph{h}_i\right) P\left(\emph{h}_i\right) d \emph{h}_i .
	\end{aligned}
\end{equation}

According to the model in \eqref{eq:009}, it is hard to determine  $\emph{\textbf{g}}_{\emph{c}}$ due to no analytical expression of $P\left(\emph{g}_{\emph{c},i}\right)$. Fortunately, the prior model of $P(\emph{\textbf{g}}_{\emph{c}}, \emph{\textbf{h}})$ can be exploited to against this difficulty, where $P(\emph{\textbf{g}}_{\emph{c}}, \emph{\textbf{h}})$ is inserted into the Eq.~\eqref{eq:007}, we have
\begin{equation}\label{eq:010}
	\begin{aligned}
		\left(\emph{\textbf{g}}_{\emph{s}}, \emph{\textbf{g}}_{\emph{c}}, \emph{\textbf{h}}, \gamma_{\emph{e}}\right)= & \underset{\emph{\textbf{g}}_{\emph{s}}, \emph{\textbf{g}}_{\emph{c}}, \emph{\textbf{h}}, \gamma_{\emph{e}}}{\operatorname{argmax}} \log P\left(\boldsymbol{r} \mid \emph{\textbf{g}}_{\emph{s}}, \emph{\textbf{g}}_{\emph{c}}, \gamma_{\emph{e}}\right) \\
		+\log P(\emph{\textbf{h}}) &+\log P(\emph{\textbf{g}}_{\emph{c}} \mid \emph{\textbf{h}})+\log P(\emph{\textbf{g}}_{\emph{s}}) .
	\end{aligned}
\end{equation}

In light of these, the  noninformative Jeffrey's prior $P\left(\emph{h}_i\right)=\frac{1}{\emph{h}_i}$ is considered to the MAP estimator of Eq.~\eqref{eq:010}, which  is given by
\begin{equation} \label{eq:010x}
	\begin{aligned}
		\left(\emph{\textbf{g}}_{\emph{s}}, \emph{\textbf{g}}_{\emph{c}}, \emph{\textbf{h}}, \gamma_{\emph{e}}\right)= & \underset{\emph{\textbf{g}}_{\emph{s}}, \emph{\textbf{g}}_{\emph{c}}, \emph{\textbf{h}}, \gamma_{\emph{e}}}{\operatorname{argmin}} \frac{1}{2 \gamma_{\emph{e}}^2}\|\boldsymbol{r}-\emph{\textbf{S}}\emph{\textbf{g}}_{\emph{s}}- \emph{\textbf{C}}\emph{\textbf{g}}_{\emph{c}}\|_2^2 \\
		&+\sqrt{2} \sum_i \frac{\left|\emph{g}_{c,i}\right|}{\emph{h}_i} +2 \sum_i \log \emph{h}_i\\
		&+\eta J(\emph{\textbf{g}}_{\emph{s}})+T\log \gamma_{\emph{e}}.
	\end{aligned}
\end{equation}

Considering the  relation of the $\emph{g}_{\emph{c},i}=\beta_i\emph{h}_{i}$, we have $\emph{\textbf{g}}_{c}=\mathrm{diag}\{\bm{\beta}\}\emph{\textbf{h}}$, where $\bm{\beta}=\{\beta_1,\beta_2,...,\beta_N\} \in \mathbb{R}^{N_b \times N_b}$. It follows that the Jeffrey's prior $P\left(\emph{h}_i\right)=\frac{1}{\emph{h}_i}$ is numerical unstable. To this end, a small constant $\epsilon$ is introduced into $P\left(\emph{h}_i\right)$, as $P\left(\emph{h}_i\right)=\frac{1}{\emph{h}_i+\epsilon}$. Thus, Eq.~\eqref{eq:010x} can be expressed as
\begin{equation} \label{eq:011}
	\begin{aligned}
		\left(\emph{\textbf{g}}_{\emph{s}}, \bm{\beta}, \emph{\textbf{h}}, \gamma_{\emph{e}}\right)= & \underset{\emph{\textbf{g}}_{\emph{s}}, \emph{\textbf{g}}_{\emph{c}}, \emph{\textbf{h}}, \gamma_{\emph{e}}}{\operatorname{argmin}} \frac{1}{2 \gamma_{\emph{e}}^2}\|\boldsymbol{r}-\emph{\textbf{S}}\emph{\textbf{g}}_{\emph{s}}-\emph{\textbf{C}}\mathrm{diag}\{\bm{\beta}\}\emph{\textbf{h}}\|_2^2 \\
		& +\sqrt{2} \sum_i\left|\beta_i\right|+2 \sum_i \log \left(\emph{h}_i+\epsilon\right)\\
		&+\eta J(\emph{\textbf{g}}_{\emph{s}})+T\log \gamma_{\emph{e}}.
	\end{aligned}
\end{equation}

It follows that the subproblem  with respect to channel $\emph{\textbf{g}}_{\emph{c}}$ can be reformulated as the joint estimation of $\bm{\beta}$ and $\emph{\textbf{h}}$. In addition, how to choose the prior model $P(\emph{\textbf{g}}_{\emph{s}})$ is also  important. In this paper, we develop weighted MCP optimization method to tackle this issue.

\section{Optimization Algorithm}\label{section4}
This section develop an alternatively optimization approach to solve the problem \eqref{eq:011}. For an initial estimate of $\bm{r}$, the alternation scheme is exploited to optimize $\emph{\textbf{h}}$ and $\bm{\beta}$ for the outliers,  and update $\emph{\textbf{g}}_{s}$ with fixed $\emph{\textbf{g}}_{c}=\mathrm{diag}\{\bm{\beta}\}\emph{\textbf{h}}$. To facilitate the optimization, the
noise variance $\gamma_{\emph{e}}$ is also integrated to optimize it by fixing the channels $\emph{\textbf{g}}_{s}$ and $\emph{\textbf{g}}_{c}$, iteratively.

\subsection{Optimization of the $\textbf{h}$-subproblem}
For given $\emph{\textbf{g}}_{s}, \bm{\beta}$ and $\gamma_{\emph{e}}$, the update for the positive hidden variables $\emph{h}_i$ is obtained as a solution by minimizing
\begin{equation} \label{eq:012}
	\begin{aligned}
		\min\limits_{\emph{\textbf{h}}}&\|\boldsymbol{r}-\emph{\textbf{S}}\emph{\textbf{g}}_{\emph{s}}-\emph{\textbf{C}}\mathrm{diag}\{\bm{\beta}\}\emph{\textbf{h}}\|_2^2+4 \gamma_{\emph{e}}^2 \sum_i \log \left(\emph{h}_{\emph{i}}+\epsilon\right),\\
		& s.t.\quad\emph{h}_i \geq 0 .
	\end{aligned}
\end{equation}

Let $\emph{\textbf{S}}=[\emph{\textbf{s}}_1^T;\emph{\textbf{s}}_2^T;...;\emph{\textbf{s}}_N^T]$ denotes the matrix formed by the set of vector. Then, the $\textbf{\emph{h}}$-subproblem \eqref{eq:012} can be rewritten as
\begin{equation} \label{eq:013}
	\begin{aligned}
		&\underset{\emph{h}_i}{\operatorname{min}} \sum_i\left(\emph{r}_{i}-\emph{\textbf{s}}_{\emph{i}}^T\emph{\textbf{g}}_{\emph{s}}-\varphi_i\emph{h}_i\right)^2+4 \gamma_{\emph{e}}^2 \sum_{i} \log \left(\emph{h}_{i}+\epsilon\right), \\
		&\quad s.t . \quad \emph{h}_i \geq 0,
	\end{aligned}
\end{equation}
where $\varphi_i=c_{\emph{i,i}}\beta_{\emph{i}}$ and $c_{\emph{i,i}}$ denotes the diagonal element-wise of the $\emph{\textbf{C}}$.

Accordingly, each $\emph{h}_i$ is independently determined  by optimizing  the following scalar optimization  problem
\begin{equation} \label{eq:014}
	\begin{aligned}
		\min\limits_{\emph{h}_i} &\left\{Q(h_i)=a_i\emph{h}_i^2+b_i\emph{h}_i+4\gamma_{\emph{e}}^2\log \left(\emph{h}_i+\epsilon\right)\right\},\\
		&s.t.\; \emph{h}_i \geq 0,
	\end{aligned}
\end{equation}
where $b_i=2\varphi_{\emph{i}}\left(\emph{\textbf{s}}_{\emph{i}}^T\emph{\textbf{g}}_{s}-\emph{r}_i\right)$ and $a_i=\varphi_i^2$. The closed-form solution can be easily derived via taking $\frac{f\left(\emph{h}_i\right)}{d \emph{h}_i}=0$. It shows that the solution of the problem \eqref{eq:014}  can be derived
$$
\emph{h}_i= \begin{cases}0, & \text {if  } \frac{\left(2 a_i \epsilon+b_i\right)^2}{16 a_i^2} -\frac{b_i \epsilon+4\gamma_{\emph{e}}^2}{2 a_i} < 0 \\ \emph{u}_i, & \text {otherwise},\end{cases}
$$
where $\emph{u}_i=\operatorname{argmin}_{\emph{h}_i}\left\{Q(0), Q\left(\emph{h}_{i, 1}\right), Q\left(\emph{h}_{i, 2}\right)\right\}$,  $\emph{h}_{i, 1}$ and $\emph{h}_{i,2}$ denotes two stationary points of $Q\left(\emph{h}_i\right)$, i.e.,
$$
\begin{aligned}
	& \emph{h}_{i, 1}=-\frac{2 a_i \epsilon+b_i}{4 a_i}+\sqrt{\frac{\left(2 a_i \epsilon+b_i\right)^2}{16 a_i^2}-\frac{b_1 \epsilon+4\gamma_{\emph{e}}^2}{2 a_i}}, \\
	& \emph{h}_{i, 2}=-\frac{2 a_i \epsilon+b_i}{4 a_i}-\sqrt{\frac{\left(2 a_i \epsilon+b_i\right)^2}{16 a_i^2}-\frac{b_i \epsilon+4\gamma_{\emph{e}}^2}{2 a_i} .}
\end{aligned}
$$

\subsection{Optimization of the $\beta$-subproblem}

For fixed $\emph{\textbf{g}}_{s}, \emph{\textbf{h}}$ and $\gamma_{\emph{e}}$, the Laplaican coefficients $\bm{\beta}$ is derived by optimizing
\begin{equation} \label{eq:017}
	\begin{aligned}
		\bm{\beta}=\underset{\bm{\beta}}{\operatorname{argmin}} \sum_{i}\left(\emph{r}_{i}-\emph{\textbf{s}}_{\emph{i}}^T\emph{\textbf{g}}_{s}-\varphi_i\emph{h}_i\right)^2+2 \sqrt{2}\gamma_{\emph{e}}^2 \sum_{i}\left|\beta_i\right| .
	\end{aligned}
\end{equation}

Based on the problem \eqref{eq:017}, each $\beta_i$ is updated by dealing with the following  problem, i.e.,
\begin{equation} \label{eq:018}
	\begin{aligned}
		\beta_i=\underset{\beta_i}{\operatorname{argmin}}\left(\emph{r}_{i}-\emph{\textbf{s}}_{\emph{i}}^T\emph{\textbf{g}}_{s}-\varphi_i\emph{h}_i\right)^2+2 \sqrt{2} \gamma_{\emph{e}}^2\left|\beta_i\right|.
	\end{aligned}
\end{equation}

In accordance with the \eqref{eq:018}, a closed-form solution is obtained, which can be generated using the soft-thresholding $\mathcal{S}_{\varsigma}(\cdot)$ as
\begin{equation} \label{eq:019}
	\begin{aligned}
		\beta_i=\mathcal{S}_{\varsigma}((\emph{r}_{i}-\emph{\textbf{s}}_{\emph{i}}^T\emph{\textbf{g}}_{s})/(\emph{h}_i+\epsilon)),
	\end{aligned}
\end{equation}
where the threshold $\varsigma=\frac{2\sqrt{2}\gamma_{\emph{e}}}{\emph{h}_i+\epsilon}$ and $\epsilon$ is considered for numerical stability \cite{8002663}.

\subsection{Optimization of the $\textbf{g}_{s}$-subproblem}
In this subsection, the MCP is employed to formulate the proposed $\textbf{\emph{g}}_{s}$-subproblem, whose aim is to update
the following problem
\begin{equation} \label{eq:020}
	\begin{aligned}
		\underset{\emph{\textbf{g}}_{s}}{\operatorname{min}}\left\{F(\emph{\textbf{g}}_{s})=\sum_{i}\left(\emph{r}_{i}-\emph{\textbf{s}}_{i}^T\emph{\textbf{g}}_{s}-\varphi_i\emph{h}_i\right)^2+\eta J(\emph{\textbf{g}}_{\emph{s}})\right\}.
	\end{aligned}
\end{equation}

To promote the convexity of the objective function in \eqref{eq:020}, the MCP regularization is employed to estimate the sensing channel, which can be defined as
\begin{equation}
	\begin{aligned}
		\varphi(x_i)= \begin{cases}|x_i|-\frac{\zeta}{2}x_i^2, & |x_i|<\frac{1}{\zeta} \\ \frac{1}{2\zeta}, & |x_i|\geq\frac{1}{\zeta}.\end{cases}
	\end{aligned}
\end{equation}

Furthermore, the MCP penalty can be expressed as a Moreau envelope as follows
\begin{equation} \label{eq:020xx}
	\begin{aligned}
		J_{\mathrm{MCP}}(\emph{\textbf{g}}_{\emph{s}})=\|\emph{\textbf{g}}_{\emph{s}}\|_1-\min _{\emph{\textbf{v}}}\left\{\|\emph{\textbf{v}}\|_1+\frac{\zeta}{2}\|\emph{\textbf{g}}_{\emph{s}}-\emph{\textbf{v}}\|_2^2\right\},
	\end{aligned}
\end{equation}
where the parameter $\zeta$ controls the nonconvexity of the MCP strategy. By taking the MCP strategy  in \eqref{eq:021}, the cost function in \eqref{eq:020} is derived by

\begin{equation} \label{eq:020x}
	\begin{aligned}
		F(\emph{\textbf{g}}_{\emph{s}})=&\|\boldsymbol{r}-\emph{\textbf{S}}\emph{\textbf{g}}_{\emph{s}}-\emph{\textbf{C}}\mathrm{diag}\{\bm{\beta}\}\emph{\textbf{h}}\|_2^2+\lambda\|\emph{\textbf{g}}_{\emph{s}}\|_1\\
		&-\lambda\min _{\emph{\textbf{v}}}\left\{\|\emph{\textbf{v}}\|_1+\frac{\zeta}{2\lambda}\|\emph{\textbf{g}}_{\emph{s}}-\emph{\textbf{v}}\|_2^2\right\}\\
		=&\emph{\textbf{g}}_{\emph{s}}^T(\emph{\textbf{S}}^T\emph{\textbf{S}}-\lambda\zeta\emph{\textbf{I}})\emph{\textbf{g}}_{\emph{s}}+\lambda\|\emph{\textbf{g}}_{\emph{s}}\| \\
		&+\max _{\emph{\textbf{v}}}\{\mathcal{W}(\emph{\textbf{g}}_{\emph{s}}, \emph{\textbf{v}})\},
	\end{aligned}
\end{equation}
where  $\mathcal{W}(\cdot,\cdot)$ denotes the affine function of sensing channel $\emph{\textbf{g}}_{\emph{s}}$, that satifies (see  \cite{20232017} , Prop. 8.14). To
maintain the convexity of the function $\emph{F}$, $\emph{\textbf{S}}^T\emph{\textbf{S}}-\lambda\zeta\emph{\textbf{I}}\succeq0$ should be satified. When the tunable $\mathcal{W}$-factor
wavelet transform is considered, the matrix $\emph{\textbf{S}}^T\emph{\textbf{S}}$ is singular that may lead to parameter  $\zeta$ as $\zeta = 0$. Yet, when $\zeta = 0$, the penalty in \eqref{eq:020xx} reduces to the $L_1$ norm.

To avoid the presence of spurious local minima in the cost function \eqref{eq:020}, a  generalized minimax-concave penalty is proposed, which can
be represented as
\begin{equation} \label{eq:021}
	\begin{aligned}
		J_{\mathrm{MCP}}(\emph{\textbf{g}}_{\emph{s}})=\|\emph{\textbf{g}}_{\emph{s}}\|_1-\min _{\emph{\textbf{v}}}\left\{\|\emph{\textbf{v}}\|_1+\frac{\zeta}{2\lambda}\|\emph{\textbf{S}}(\emph{\textbf{g}}_{\emph{s}}-\emph{\textbf{v}})\|_2^2\right\},
	\end{aligned}
\end{equation}
where the parameter $\zeta$ controls the nonconvexity of the MCP strategy. By taking the MCP strategy  in \eqref{eq:021}, the cost function in \eqref{eq:020} is derived by
\begin{equation}
	\begin{aligned}
		F(\emph{\textbf{g}}_{\emph{s}})= & \frac{1}{2}\|\boldsymbol{r}-\emph{\textbf{S}}\emph{\textbf{g}}_{\emph{s}}-\emph{\textbf{C}}\mathrm{diag}\{\bm{\beta}\}\emph{\textbf{h}}\|_2^2+\lambda\|\emph{\textbf{g}}_{\emph{s}}\|_1 \\
		& -\min _{\emph{\textbf{v}}}\left\{\lambda\|\emph{\textbf{v}}\|_1+\frac{\zeta}{2}\|\emph{\textbf{S}}(\emph{\textbf{g}}_{\emph{s}}-\emph{\textbf{v}})\|_2^2\right\} \\
		= & \max _{\emph{\textbf{v}}}\left\{\frac{1}{2}\|\boldsymbol{r}-\emph{\textbf{S}}\emph{\textbf{g}}_{\emph{s}}-\emph{\textbf{C}}\mathrm{diag}\{\bm{\beta}\}\emph{\textbf{h}}\|_2^2+\lambda\|\emph{\textbf{g}}_{\emph{s}}\|_1\right. \\
		& \left.-\lambda\|\emph{\textbf{v}}\|_1-\frac{\zeta}{2}\|\emph{\textbf{S}}(\emph{\textbf{g}}_{\emph{s}}-\emph{\textbf{v}})\|_2^2\right\} \\
		= & \max _{\emph{\textbf{v}}}\left\{\frac{1}{2}(1-\zeta)\|\emph{\textbf{S}}\emph{\textbf{g}}_{\emph{s}}\|_2+\lambda\|\emph{\textbf{g}}_{\emph{s}}\|_1+\mathcal{W}(\emph{\textbf{g}}_{\emph{s}}, \emph{\textbf{v}})\right\} \\
		= & \frac{1}{2}(1-\zeta)\|\emph{\textbf{S}}\emph{\textbf{g}}_{\emph{s}}\|_2+\lambda\|\emph{\textbf{g}}_{\emph{s}}\|_1+\max _{\emph{\textbf{v}}}\{\mathcal{W}(\emph{\textbf{g}}_{\emph{s}}, \emph{\textbf{v}})\},
	\end{aligned}
\end{equation}
where  $\mathcal{W}(\cdot,\cdot)$ denotes the affine function of sensing channel $\emph{\textbf{g}}_{\emph{s}}$, that satifies (see  \cite{20232017} , Prop. 8.14). Therefore, the  objective function $F(\emph{\textbf{g}}_{\emph{s}})$ is convex if the parameter $\zeta$ satisfies  $0<\zeta\leqslant 1$.

Since the MCP penalty cannot directly obtain a closed-form formula, an iterative algorithm is exploited to  evaluate the feasible solution, which includes  a double nested loop: an inner loop is employed to find the optimal solution $\emph{\textbf{v}}^{\text {opt }}$, while the optimal solution $\emph{\textbf{g}}_{\emph{s}}^{\text {opt }}$ can be iteratively obtained by taking an outer loop.
Fortunately, $F(\emph{\textbf{g}}_{\emph{s}})$ can be modeled as a saddle-point convex form in $\emph{\textbf{g}}_{\emph{s}}$ and concave in $\emph{\textbf{v}}$, the corresponding solution can be obtained by employing proximal algorithm.
Based on the aforementioned analysis, the MCP-regularized least squares minimization problem is formulated as
\begin{equation}
	\begin{aligned}
		\left(\emph{\textbf{g}}_{\emph{s}}^{\text {opt }}, \emph{\textbf{v}}^{\text {opt }}\right)=\arg \min _{\emph{\textbf{g}}_{\emph{s}}} \max _{\emph{\textbf{v}}} F(\emph{\textbf{g}}_{\emph{s}}, \emph{\textbf{v}}),
	\end{aligned}
\end{equation}
where
$$
\begin{aligned}
	F(\emph{\textbf{g}}_{\emph{s}}, \emph{\textbf{v}})=&\frac{1}{2}\|\boldsymbol{r}-\emph{\textbf{S}}\emph{\textbf{g}}_{\emph{s}}-\emph{\textbf{C}}\mathrm{diag}\{\bm{\beta}\}\emph{\textbf{h}}\|_2^2+\lambda\|\emph{\textbf{g}}_{\emph{s}}\|_1\\
	&\qquad-\lambda\|\emph{\textbf{v}}\|_1-\frac{\zeta}{2}\|\emph{\textbf{S}}(\emph{\textbf{g}}_{\emph{s}}-\emph{\textbf{v}})\|_2^2,
\end{aligned}
$$
with $0<\zeta\leq1$. The preceding sequence is summarized in Algorithm \ref{alg1}, we apply the forward-backward splitting algorithm to solve the  saddle-point problem, where the $\mu$ satisfies $\mu\leq\frac{2}{\max\{1, \zeta/(1-\zeta)\}\|\emph{\textbf{S}}\|_2^{-2}}$ to guarantee the convergence.

\begin{algorithm}
	\caption{Iterative algorithm for solving problem \eqref{eq:020}.}
	\label{alg1}
	\begin{algorithmic}[1]
		\STATE \textbf{Input}:  initial values $\emph{\textbf{g}}_{\emph{s}}$, $\emph{\textbf{v}}$
		\STATE \qquad$0<\mu<2/\max\{1, \gamma/(1-\gamma)\}\|\bm{\Phi}\|_2^{-2}$
		\STATE \quad  \textbf{repeat}
		\STATE \qquad\; Update $\tilde{\emph{\textbf{g}}}_{\emph{s}}^{(\emph{k})}=\bm{\nu}^{(\emph{k})}-\lambda\boldsymbol{\Phi}^T(\boldsymbol{\Phi}\emph{\textbf{g}}_{\emph{s}}+$
		\STATE \qquad\quad$\boldsymbol{\Phi}\mathrm{diag}\{\bm{\beta}\}\emph{\textbf{h}}-\boldsymbol{r})+\lambda\gamma\boldsymbol{\Phi}^T\boldsymbol{\Phi}(\emph{\textbf{g}}_{\emph{s}}^{(\emph{k})}-\emph{\textbf{v}}^{(\emph{k}}))$;
		\STATE \qquad\; Update $\tilde{\emph{\textbf{v}}}^{(\emph{k})}=\emph{\textbf{v}}^{(\emph{k})}+\lambda\gamma\boldsymbol{\Phi}^T\boldsymbol{\Phi}(\emph{\textbf{g}}_{\emph{s}}^{(\emph{k})}-\emph{\textbf{v}}^{(\emph{k}}))$;
		\STATE \qquad\; Update $\emph{\textbf{g}}_{\emph{s}}^{(\emph{k}+1)}=\mathcal{S}_{\lambda\mu}(\tilde{\emph{\textbf{g}}}_{\emph{s}}^{(\emph{k})})$;
		\STATE \qquad\; Update $\emph{\textbf{v}}^{(\emph{k}+1)}=\mathcal{S}_{\lambda\mu}(\tilde{\emph{\textbf{v}}}^{(\emph{k})})$;
		\STATE \quad \textbf{Until} The obtained solution is feasible;
		\STATE \textbf{Output} $\emph{\textbf{g}}_{\emph{s}}^{(\emph{i}+1)}$
	\end{algorithmic}
\end{algorithm}

\subsection{Optimization of the $\gamma_{\emph{\emph{e}}}$-subproblem}
For any updated $\emph{\textbf{g}}_{\emph{s}}, \bm{\beta}$ and $\emph{\textbf{h}}$, the noise variance $\gamma_{\emph{e}}$ can
be estimated by solving
\begin{equation} \label{eq:022}
	\begin{aligned}
		\underset{\gamma_{\emph{e}}}{\operatorname{min}}\left\{ \frac{1}{2 \gamma_{\emph{e}}^2}\|\boldsymbol{r}-\emph{\textbf{S}}\emph{\textbf{g}}_{\emph{s}}- \emph{\textbf{C}}\emph{\textbf{g}}_{\emph{c}}\|_2^2 +T\log \gamma_{\emph{e}}
		\right\}.
	\end{aligned}
\end{equation}

By taking the $\frac{F(\gamma_{\emph{e}})}{d\gamma_{\emph{e}}}=0$, the closed-form solution can be obtained as
\begin{equation} \label{eq:023}
	\begin{aligned}
		\gamma_{\emph{e}}= \sqrt{\|\boldsymbol{r}-\emph{\textbf{S}}\emph{\textbf{g}}_{\emph{s}}-\emph{\textbf{C}}\emph{\textbf{g}}_{\emph{c}}\|_2^2/T}.
	\end{aligned}
\end{equation}

The preceding sequence of the S$\&$C channels, $\{\emph{\textbf{g}}_{\emph{s}}, \emph{\textbf{g}}_{\emph{c}}\}$ and the noise variance $\gamma_{\emph{\emph{e}}}$ is listed in Algorithm \ref{alg2}.

\begin{algorithm}
	\caption{Overall alternating for solving problem \eqref{eq:011}.}
	\label{alg2}
	\begin{algorithmic}[1]
		\STATE \textbf{Input}:  initial values $\emph{\textbf{g}}_{\emph{s}}$, $\emph{\textbf{v}}$
		\STATE \quad  \textbf{repeat}
		\STATE \qquad\; Update $\textbf{\emph{h}}$ for given  $\emph{\textbf{g}}_{s}, \bm{\beta}$ and $\gamma_{\emph{e}}$ according to \eqref{eq:012};
		\STATE \qquad\; Update $\bm{\beta}$ for given  $\emph{\textbf{g}}_{s}, \textbf{\emph{h}}$ and $\gamma_{\emph{e}}$ according to \eqref{eq:017};
		\STATE \qquad\; Update $\emph{\textbf{g}}_{s}$ using Algorithm 1;
		\STATE \qquad\; Update $\gamma_{\emph{e}}$ for given  $\bm{\beta}, \textbf{\emph{h}}$ and $\emph{\textbf{g}}_{s}$ according to \eqref{eq:022};
		\STATE \quad \textbf{Until} The obtained solution is feasible;
		\STATE \textbf{Output} $\textbf{\emph{h}}$, $\emph{\textbf{g}}_{\emph{s}}$
	\end{algorithmic}
\end{algorithm}

\section{simulation results}
\label{SIMULATION-RESULTS}

We provide the simulation results of the proposed S$\&$C estimation approach in comparison with some benchmarks. the proposed S$\&$C approach with generalized minimax-concave penalty is evaluated under
various SNR cases. According to 3G PP standard, the 3.5 GHz frequency band is operated on the ISAC system. The pilot symbol
duration is $T = 0.5\mu s$ and $L =  64$  unless further specified. Similar to the assumptions in \cite{10473705}, we consider a crossroad scenario in the simulation where vehicles are dropped in the crossroad and a BS is located at the center.

\subsection{The NMSE Performance versus SNR}
\begin{figure}[!t]
	\centering
	\includegraphics[width=0.5\textwidth]{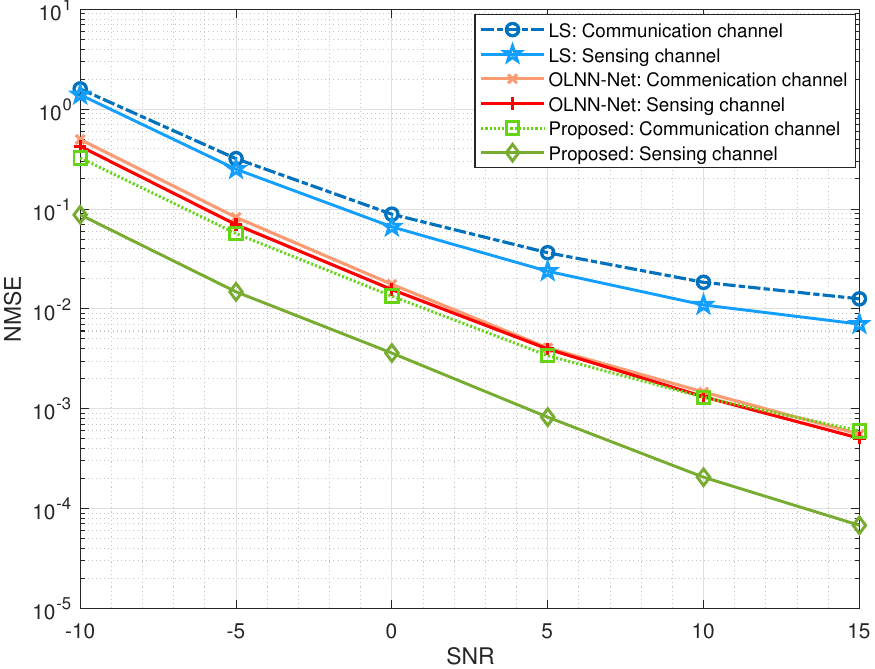}
	\caption{NMSE of ISAC channel estimation vs SNR.}\label{fig3}
\end{figure}
The NMSE performance of the ISAC channels estimation is compared with the LS and  deep learning-based benchmark in \cite{9622178}. Specifically, Figure \ref{fig3} illustrates the comparative performance of various algorithms in estimating the sensing and communication channels, denoted as $\emph{\textbf{g}}_s$ (sensing channel) and $\emph{\textbf{g}}_c$ (communication channel). When comparing the proposed estimation method to the LS algorithm, a notable improvement is observed. For the sensing channel $\emph{\textbf{g}}_s$, the proposed method achieves an NMSE improvement of $10^1$, demonstrating a significant enhancement in estimation accuracy over the LS scheme. Additionally, for the communication channel $\emph{\textbf{g}}_c$, the proposed scheme provides approximately a 3 dB improvement in  SNR over the LS method, indicating better performance in terms of communication channel estimation. Further analysis leads to the conclusion that the NMSE performance achieved by the proposed method is on par with that of the deep learning-based benchmark, especially when identical element-grouping sizes are used in the comparison. This favorable outcome is attributed to the simplicity of the sensing model used in the proposed method, which is inherently less complex than the communication model. As a result, the proposed method is particularly well-suited for application in large-scale RIS arrays, where managing both sensing and communication channels is crucial.

The reduced complexity of the sensing model not only facilitates more efficient channel estimation but also makes the method more practical for massive RIS deployments, where computational resources and power efficiency are critical concerns. This simplicity allows for robust performance in large-scale systems while maintaining competitive accuracy levels compared to more complex models like those based on deep learning.

\begin{figure}[!t]
	\centering
	\includegraphics[width=0.5\textwidth]{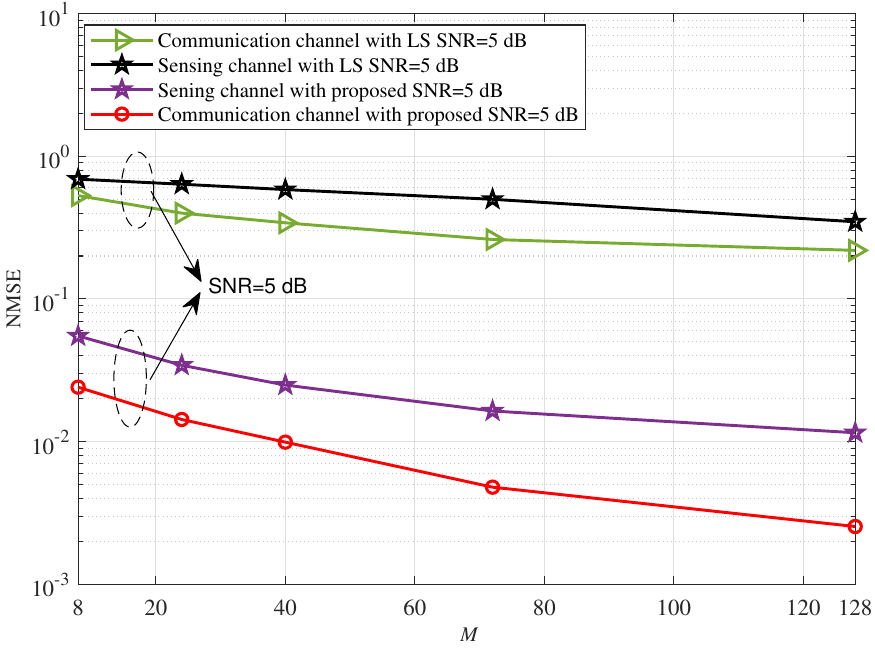}
	\caption{NMSE of ISAC channel estimation vs $M$.}\label{fig4}
\end{figure}

\subsection{The Performance versus Transmit Antennas}
Figure \ref{fig4} depicts 
the impact of different $M$ value on the estimation performance. We fix the element of RIS $L=64$ and the SNR value is the same as in Figure \ref{fig3}. Similar to the findings of Figure \ref{fig3}, the proposed approach provides significant NMSE performance enhancement compared to the LS baseline scheme to verify the estimation performance for various $M$ values. This is due to
the fact that the NMSE slightly reduces with the increase of $M$. Moreover, it is obviously that our proposed approach has better and more stable performance than other algorithms. This may lie in the fact that the proposed approach possesses a remarkable CSI features to enhance the estimation result.

\subsection{The Performance versus Element-grouping Size}
\begin{figure}
	\centering
	\includegraphics[width=0.495\textwidth]{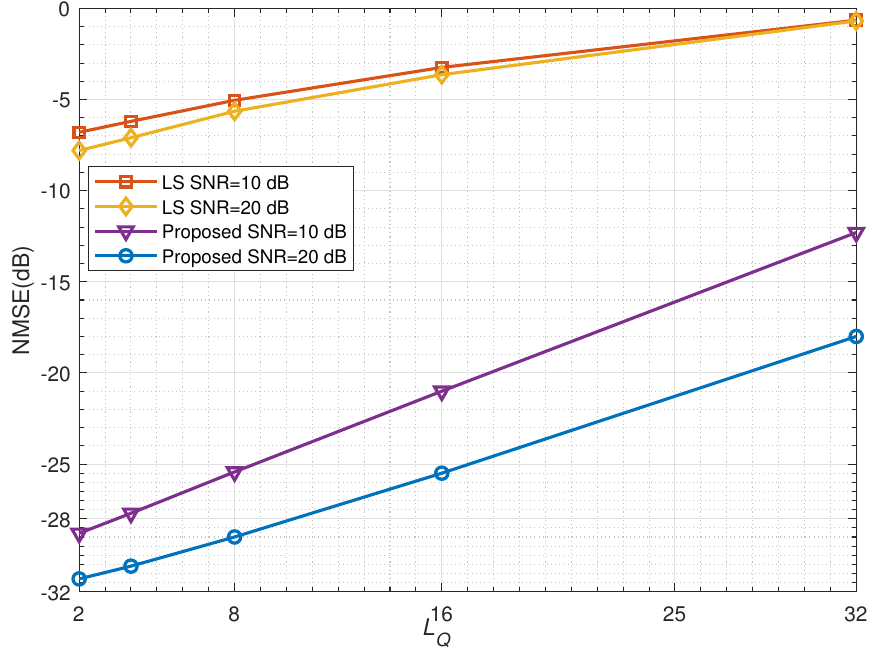}
	\caption{NMSE of ISAC channel estimation vs $L_Q$.}\label{fig5}
\end{figure}

Figure \ref{fig5} demonstrates the NMSE performance versus element-grouping size $L_Q$. For fair comparison, we consider the performance of two different benchmarks with SNR = 10 dB and 20 dB. The estimation accuracy of LS baseline scheme is poor due to the intragroup interference. The proposed scheme with  element-grouping strategy achieve higher gains compared to LS baseline approach under different $L_Q$, which verifies the powerful extrapolation ability of the proposed approach. Moreover, due to the refinement of partial channels, the proposed MCP-based approach slightly outperforms LS baseline scheme, which verifies the effectiveness of the proposed MCP-based method.

\subsection{The Performance versus Passive Reflecting Elements}
\begin{figure}[!ht]
	\includegraphics[width=0.495\textwidth]{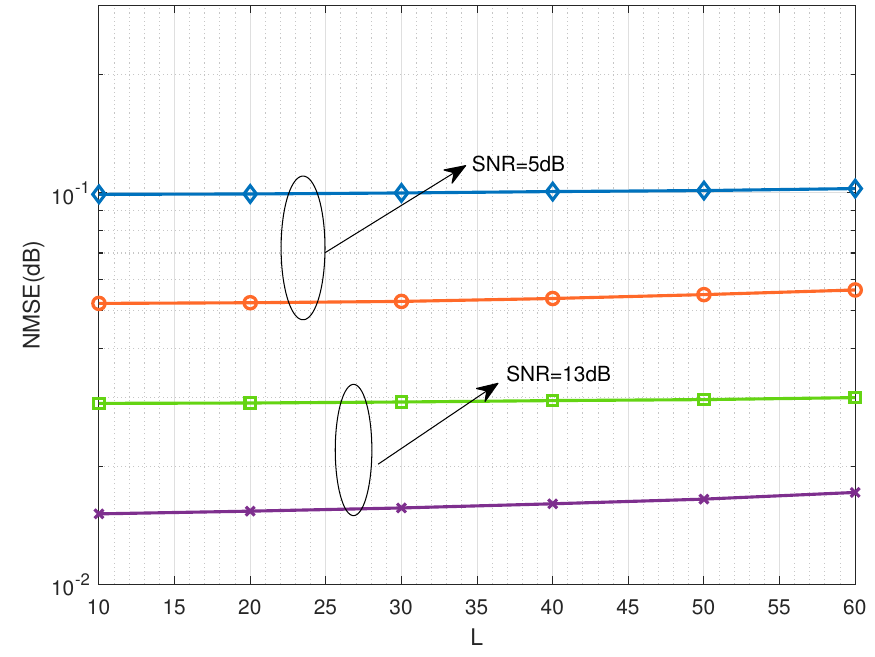}
	\caption{NMSE of ISAC channel estimation vs $L_Q$.}\label{fig6}
\end{figure}

Since $L$ influences the dimension of the communication channel $\emph{\textbf{g}}_{\emph{c}}$, we assess the impact of varying $L$ on estimation performance in Figure \ref{fig6}. The SNR  is fixed to 5dB and 13 dB for LS and the proposed method. Obviously, the proposed approach demonstrates significant performance enhancement across different $L$ values and SNR conditions compared to the benchmark scheme. However, it is worth noting a slight increase in NMSE as $L$ increases. This increase can be attributed to the greater complexity involved in mapping the pair $\{\emph{\textbf{g}}_{\emph{s}},\emph{\textbf{g}}_{\emph{c}}\}$.  Estimation becomes more challenging for the LS method as channel dimensions expand, thereby affecting the accuracy of the estimation.

\begin{figure}[!ht]
	\centering
	\includegraphics[width=0.495\textwidth]{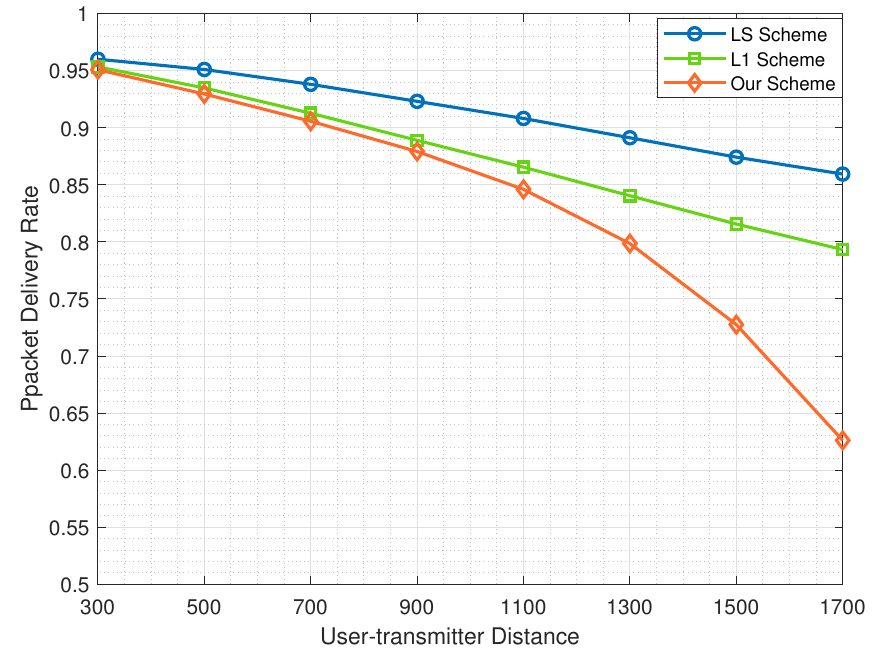}
	\caption{Packet delivery rate vs the user-transmitter distance.}\label{fig7}
\end{figure}

To verify the practicability of IoV communication, packet delivery rate is considered in this subsection. Figure \ref{fig7} illustrates that as the vehicle-transmitter distance increases, the packet delivery rates for vehicle data in three alternative methods exhibit downward trends. This decline is attributed to the reduction in effective channel links as the vehicle-transmitter distance increases, which compromises the adequacy of existing channels for vehicle data communication and, consequently, reduces the packet delivery rate. After being compared to other schemes, it can be seen that our method significantly outperform the corresponding benchmarks. These findings underscore the importance of ISAC channel estimation in enhancing the quality of the RIS-ISAC network. the proposed design has much lower computational complexity than that of the conventional LS one.

Finally the complexity of our proposed algorithm is compared with that of the conventional LS algorithm. LS estimation complexity is dominated by a matrix inverse, i.e., $\mathcal{O}(M^3)$, ignoring the costs of estimating second order channel and noise statistics. Considering the number of RF chains is limited aiming to save the cost and energy, which leads to a limited $N_{\emph{RF}}$ satisfying $N_{\emph{RF}}<M$. Thus, for each UE, the total computational complexity of our proposed method is of  $\mathcal{O}(N_{\emph{RF}}T+LM)$, where $T$ is the length of the transmission block. This observation confirms the remark that the proposed design has much lower computational complexity than that of the conventional  one. Specifically, the method minimizes the number of operations needed for channel estimation by leveraging the inherent sparsity in the RIS-assisted ISAC channel and adopting a weighted minimax-concave penalty (MCP) approach. This significantly reduces the training overhead and computational burden compared to conventional methods.

\section{Conclusion}\label{section6}
This paper investigated a novel channel estimation approach for RIS-assisted ISAC systems, addressing the limitations of existing methods that rely on extensive pilot sequences. Unlike existing schemes that depend on extensive pilot sequences, the proposed scheme jointly employ the MCP and the element-grouping strategy, which can significant improve the sensing and communication channel estimation accuracy. Numerical results have shown that compared to the benchmark scheme, the proposed approach possesses a substantial NMSE performance improvement under different SNR conditions and channel dimensions.

\section*{ACKNOWLEDGEMENT}
\label{ACKNOWLEDGEMENT}
This work has been supported in part by the National Natural Science Foundation of China under Grant 62001171, in part by the Natural Science Foundation of Guangdong Province under Grant 2024A1515011172, in part by the Henan Science and Technology Research and Development Program Joint Fund under Grant 235200810049.

\bibliography{myref}

\begin{thebibliography}{32}
\providecommand{\natexlab}[1]{#1}
\providecommand{\url}[1]{#1}
\expandafter\ifx\csname urlstyle\endcsname\relax\else
  \urlstyle{same}\fi
\expandafter\ifx\csname href\endcsname\relax
  \DeclareUrlCommand\doi{\urlstyle{rm}}
  \def\eprint#1#2{#2}
\else
  \def\doi#1{\href{https://doi.org/#1}{\nolinkurl{#1}}}
  \let\eprint\href
\fi

\bibitem[Chen et~al.(2023)Chen, Ye, and Huang]{10118701}
CHEN Z, YE J, HUANG L.
\newblock A two-stage beamforming design for active ris aided dual functional
  radar and communication\allowbreak[C]//\allowbreak
2023 IEEE Wireless Communications and Networking Conference (WCNC).
\newblock 2023: 1-6.

\bibitem[Wang et~al.(2023)Wang, Mu, and Liu]{10135096}
WANG Z, MU X, LIU Y.
\newblock Near-field integrated sensing and communications\allowbreak[J].
\newblock IEEE Communications Letters, 2023, 27\allowbreak (8): 2048-2052.

\bibitem[Hua et~al.(2024)Hua, Wu, Chen, Dobre, and Swindlehurst]{10143420}
HUA M, WU Q, CHEN W, et~al.
\newblock Secure intelligent reflecting surface-aided integrated sensing and
  communication\allowbreak[J].
\newblock IEEE Transactions on Wireless Communications, 2024, 23\allowbreak
  (1): 575-591.

\bibitem[Zhang et~al.(2023)Zhang, Wang, Wu, Guan, and Liu]{10325366}
ZHANG H, WANG D, WU S, et~al.
\newblock Ustb 6g: Key technologies and metaverse applications\allowbreak[J].
\newblock IEEE Wireless Communications, 2023, 30\allowbreak (5): 112-119.

\bibitem[Lu et~al.(2023)Lu, Liu, and Hanzo]{9906898}
LU S, LIU F, HANZO L.
\newblock The degrees-of-freedom in monostatic isac channels: Nlos exploitation
  vs. reduction\allowbreak[J].
\newblock IEEE Transactions on Vehicular Technology, 2023, 72\allowbreak (2):
  2643-2648.

\bibitem[Xiao et~al.(2022)Xiao and Zeng]{9724187}
XIAO Z, ZENG Y.
\newblock Waveform design and performance analysis for full-duplex integrated
  sensing and communication\allowbreak[J].
\newblock IEEE Journal on Selected Areas in Communications, 2022, 40\allowbreak
  (6): 1823-1837.

\bibitem[Liu et~al.(2020)Liu and Shoji]{8809218}
LIU W, SHOJI Y.
\newblock Deepvm: Rnn-based vehicle mobility prediction to support intelligent
  vehicle applications\allowbreak[J].
\newblock IEEE Transactions on Industrial Informatics, 2020, 16\allowbreak (6):
  3997-4006.

\bibitem[Zhang et~al.(2023)Zhang, He, Ai, Yang, Zhang, Chen, Zhang, and
  Zhong]{10225688}
ZHANG Z, HE R, AI B, et~al.
\newblock A shared multipath components evolution model for integrated sensing
  and communication channels\allowbreak[J].
\newblock IEEE Antennas and Wireless Propagation Letters, 2023, 22\allowbreak
  (12): 2975-2978.

\bibitem[Cheng et~al.(2022)Cheng, Duan, Gao, and Yang]{9830717}
CHENG X, DUAN D, GAO S, et~al.
\newblock Integrated sensing and communications (isac) for vehicular
  communication networks (vcn)\allowbreak[J].
\newblock IEEE Internet of Things Journal, 2022, 9\allowbreak (23):
  23441-23451.

\bibitem[Lin et~al.(2024)Lin, Zheng, Vorobyov, Zhou, and Shi]{10448481}
LIN L, ZHENG H, VOROBYOV S~A, et~al.
\newblock Sensing-aided communication channel estimation with tensor-based
  moving target localization\allowbreak[C]//\allowbreak
ICASSP 2024 - 2024 IEEE International Conference on Acoustics, Speech and
  Signal Processing (ICASSP).
\newblock 2024: 8606-8610.

\bibitem[Liu et~al.(2022)Liu, Li, Liu, Lu, and Han]{9787809}
LIU Y, LI M, LIU A, et~al.
\newblock Information-theoretic limits of integrated sensing and communication
  with correlated sensing and channel states for vehicular
  networks\allowbreak[J].
\newblock IEEE Transactions on Vehicular Technology, 2022, 71\allowbreak (9):
  10161-10166.

\bibitem[Chen et~al.(2022)Chen, Tang, Zhang, Wu, Wang, So, Jin, and
  Wong]{9622178}
CHEN Z, TANG J, ZHANG X~Y, et~al.
\newblock Offset learning based channel estimation for intelligent reflecting
  surface-assisted indoor communication\allowbreak[J].
\newblock IEEE Journal of Selected Topics in Signal Processing, 2022,
  16\allowbreak (1): 41-55.

\bibitem[Wei et~al.(2023)Wei, Wu, Mishra, and Bhavani~Shankar]{10218356}
WEI T, WU L, MISHRA K~V, et~al.
\newblock Multi-irs-aided doppler-tolerant wideband dfrc system\allowbreak[J].
\newblock IEEE Transactions on Communications, 2023: 1-1.

\bibitem[Chen et~al.(2022)Chen, Tang, Zhang, So, Jin, and Wong]{9521836}
CHEN Z, TANG J, ZHANG X~Y, et~al.
\newblock Hybrid evolutionary-based sparse channel estimation for irs-assisted
  mmwave mimo systems\allowbreak[J].
\newblock IEEE Transactions on Wireless Communications, 2022, 21\allowbreak
  (3): 1586-1601.

\bibitem[Li et~al.(2023)Li, Zhou, Gong, and Liu]{10186271}
LI J, ZHOU G, GONG T, et~al.
\newblock Beamforming design for active irs-aided mimo integrated sensing and
  communication systems\allowbreak[J].
\newblock IEEE Wireless Communications Letters, 2023, 12\allowbreak (10):
  1786-1790.

\bibitem[Lu et~al.(2021)Lu, Lin, Song, Fang, Hua, and Deng]{9361184}
LU W, LIN Q, SONG N, et~al.
\newblock Target detection in intelligent reflecting surface aided distributed
  mimo radar systems\allowbreak[J].
\newblock IEEE Sensors Letters, 2021, 5\allowbreak (3): 1-4.

\bibitem[Chu et~al.(2023)Chu, Nguyen, Hoang, Pham, Phan, Hwang, and
  Dutkiewicz]{10109100}
CHU N~H, NGUYEN D~N, HOANG D~T, et~al.
\newblock Ai-enabled mm-waveform configuration for autonomous vehicles with
  integrated communication and sensing\allowbreak[J].
\newblock IEEE Internet of Things Journal, 2023, 10\allowbreak (19):
  16727-16743.

\bibitem[Xu et~al.(2024)Xu, Xia, Li, Wei, Xie, and Shi]{10286864}
XU K, XIA X, LI C, et~al.
\newblock Channel feature projection clustering based joint channel and doa
  estimation for isac massive mimo ofdm system\allowbreak[J].
\newblock IEEE Transactions on Vehicular Technology, 2024, 73\allowbreak (3):
  3678-3689.

\bibitem[Yang et~al.(2023)Yang, Wang, Huang, Aggoune, and Hao]{10078840}
YANG R, WANG C~X, HUANG J, et~al.
\newblock A novel 6g isac channel model combining forward and backward
  scattering\allowbreak[J].
\newblock IEEE Transactions on Wireless Communications, 2023, 22\allowbreak
  (11): 8050-8065.

\bibitem[Zhang et~al.(2023)Zhang, Huang, Tan, Yuan, and Liu]{10349836}
ZHANG X, HUANG H, TAN L, et~al.
\newblock Enhanced channel estimation for otfs-assisted isac in vehicular
  networks: A deep learning approach\allowbreak[C]//\allowbreak
2023 21st International Symposium on Modeling and Optimization in Mobile, Ad
  Hoc, and Wireless Networks (WiOpt).
\newblock 2023: 703-707.

\bibitem[Cai et~al.(2023)Cai, Chen, Chen, Yin, and Wang]{10128162}
CAI S, CHEN L, CHEN Y, et~al.
\newblock Pulse-based isac: Data recovery and ranging estimation for multi-path
  fading channels\allowbreak[J].
\newblock IEEE Transactions on Communications, 2023, 71\allowbreak (8):
  4819-4838.

\bibitem[Memisoglu et~al.(2023)Memisoglu, Türkmen, Ozbakis, and
  Arslan]{10082967}
MEMISOGLU E, TüRKMEN H, OZBAKIS B~A, et~al.
\newblock Csi-based noma for integrated sensing and
  communication\allowbreak[J].
\newblock IEEE Wireless Communications Letters, 2023, 12\allowbreak (6):
  1086-1090.

\bibitem[Sang et~al.(2024)Sang, Yuan, Tang, Li, Li, Jin, Cheng, and
  Cui]{9999288}
SANG J, YUAN Y, TANG W, et~al.
\newblock Coverage enhancement by deploying ris in 5g commercial mobile
  networks: Field trials\allowbreak[J].
\newblock IEEE Wireless Communications, 2024, 31\allowbreak (1): 172-180.

\bibitem[Sang et~al.(2024)Sang, Zhou, Lan, Gao, Tang, Li, Jin, Basar, Li,
  Cheng, and Cui]{10319362}
SANG J, ZHOU M, LAN J, et~al.
\newblock Multi-scenario broadband channel measurement and modeling for sub-6
  ghz ris-assisted wireless communication systems\allowbreak[J].
\newblock IEEE Transactions on Wireless Communications, 2024, 23\allowbreak
  (6): 6312-6329.

\bibitem[Liu et~al.(2023)Liu, Al-Nahhal, Dobre, and Wang]{9997576}
LIU Y, AL-NAHHAL I, DOBRE O~A, et~al.
\newblock Deep-learning channel estimation for irs-assisted integrated sensing
  and communication system\allowbreak[J].
\newblock IEEE Transactions on Vehicular Technology, 2023, 72\allowbreak (5):
  6181-6193.

\bibitem[Elsayed et~al.(2022)Elsayed, El-Banna, Dobre, Shiu, and Wang]{9736621}
ELSAYED M, EL-BANNA A~A~A, DOBRE O~A, et~al.
\newblock Hybrid-layers neural network architectures for modeling the
  self-interference in full-duplex systems\allowbreak[J].
\newblock IEEE Transactions on Vehicular Technology, 2022, 71\allowbreak (6):
  6291-6307.

\bibitem[Zhang et~al.(2021)Zhang, Zhang, Gao, Ma, and Dobre]{9530267}
ZHANG S, ZHANG S, GAO F, et~al.
\newblock Deep learning-based ris channel extrapolation with
  element-grouping\allowbreak[J].
\newblock IEEE Wireless Communications Letters, 2021, 10\allowbreak (12):
  2644-2648.

\bibitem[Yin et~al.(2022)Yin, Li, Tian, and Yu]{9621762}
YIN S, LI Y, TIAN Y, et~al.
\newblock Intelligent reflecting surface enhanced wireless communications with
  deep-learning-based channel prediction\allowbreak[J].
\newblock IEEE Transactions on Vehicular Technology, 2022, 71\allowbreak (1):
  1049-1053.

\bibitem[Box et~al.(1973)Box and Tiao]{1973Bayesian}
BOX G, TIAO G~C.
\newblock Bayesian inference in statistical analysis\allowbreak[J].
\newblock International Statistical Review, 1973, 43: 242.

\bibitem[Chen et~al.(2017)Chen, Fu, Xiang, and Rong]{8002663}
CHEN Z, FU Y, XIANG Y, et~al.
\newblock A novel iterative shrinkage algorithm for cs-mri via adaptive
  regularization\allowbreak[J].
\newblock IEEE Signal Processing Letters, 2017, 24\allowbreak (10): 1443-1447.

\bibitem[Combettes(2017)]{20232017}
COMBETTES.
\newblock Convex analysis and monotone operator theory in hilbert spaces (cms
  books in mathematics) by heinz h. bauschke;patrick l.\allowbreak[Z].
\newblock 2017.

\bibitem[Chen et~al.(2024)Chen, Zhang, So, Wong, Chae, and Wang]{10473705}
CHEN Z, ZHANG X~Y, SO D~K~C, et~al.
\newblock Federated learning driven sparse code multiple access in v2x
  communications\allowbreak[J].
\newblock IEEE Network, 2024: 1-8.

\end{thebibliography}

\newpage
~\\
\vspace{-11mm}
\biographies

\begin{CCJNLbiography}{cz}{Chen Zhen}
received the M.S. degree in software engineering form Xiamen
University in 2012 and the Ph.D. degree in electronic engineering from South China University of Technology, Guangzhou, China, in 2019. From 2020 to 2023, he was a Lead Engineer with Hong Kong Applied Science and Technology Research Institute, Hong Kong. From 2023 to 2024, he was a Post-Doctoral Research Fellow with the Department of Electrical Engineering, City University of Hong Kong, Hong Kong. He is currently a Research Fellow with the Institute of Microelectronics, University of Macau, Macau. His current research interests include large language model, integrated radar and communication, channel estimation, AI communications, and 5G/6G networks. He was the exemplary reviewer of several IEEE journals. He has also on the editorial board of several international journals, including Digital Communications and Networks, Signal Processing (Elsevier), IEEE OPEN JOURNAL OF SIGNAL PROCESSING, and IEEE OPEN
JOURNAL OF THE COMMUNICATIONS SOCIETY.
\end{CCJNLbiography}

\begin{CCJNLbiography}{ljq}{Li Jianqing}
 received the Ph.D. degree from Beijing University of Posts and Telecommunications, Beijing, China, in April 1999. From 2000 to 2002, he was a visiting professor of Information and Communications University, Daejeon, Korea. From 2002 to 2004, he was a research fellow of Nanyang Technological University, Singapore. He joined the Macau University of Science and Technology in August 2004. Currently, he is a professor. His research interests are IoT, machine learning, wireless networks, and fiber sensors. He is a senior member of the Institute of Electrical and Electronics Engineers (IEEE).
\end{CCJNLbiography}
\newpage
~\\
\begin{CCJNLbiography}{zhj}{Zhang Haijun}
is currently a Full Professor with the University of Science and Technology Beijing,
China. He was a postdoctoral research fellow with University of British Columbia (UBC), Canada. He serves as an editor for IEEE Transactions on Information Forensics and Security and IEEE Transactions on Network and Service Engineering. He is a IEEE ComSoc Distinguished Lecturer.
\end{CCJNLbiography}

\begin{CCJNLbiography}{zw}{Zhang Wei}
Senior Engineer at the professorial level of Henan Academy of Sciences, a master's supervisor, an expert honored with the special allowance from the provincial government, and the academic leader of the scientific and technological innovation team in Henan Province. He graduated from Xinjiang University with a bachelor's degree in Electronic Information, obtained a master's degree in Communication from the PLA Information Engineering University, and a doctorate degree in Optical Engineering from Henan Normal University. His current research directions are wireless communication, artificial intelligence, and navigation applications.He led his team to successively overcome a series of key technologies, such as channel amplitude-phase error correction and spatio-temporal-frequency adaptive filtering, achieving technological breakthroughs in the field of navigation anti-jamming in China. He has successively presided over and completed more than ten major scientific and technological projects, including national informatization special projects, high-quality development special projects of the Ministry of Industry and Information Technology, and key projects of the Henan Joint Fund. Currently, he serves as a reviewer for Henan Science, Journal of Terahertz Science and Electronic Information Technology, and Global Positioning System.
\end{CCJNLbiography}

\end{document}